\documentstyle[12pt]{article}
\def\nue{{\nu_e}}
\def\anue{{\bar\nu_e}}
\def\numu{{\nu_{\mu}}}
\def\anumu{{\bar\nu_{\mu}}}
\def\nutau{{\nu_{\tau}}}
\def\anutau{{\bar\nu_{\tau}}}

\begin{document}

\begin{center}
{\large \bf Effect of Flavour Oscillations on 
the Detection of Supernova Neutrinos}
\end{center}
\begin{center}
Sandhya Choubey$^1$, Debasish Majumdar$^2$ and Kamales Kar$^1$ \\
$^1$Saha Institute of Nuclear Physics, \\ 1/AF Bidhan Nagar, 
Calcutta 700 064, India \\
$^2$ Department of Physics, University of Calcutta, \\
92 Acharya Prafulla Chandra Road, Calcutta 700 009, India 
\end{center}

{\bf Abstract}:{\small Neutrinos and antineutrinos of all three flavours 
are emitted during the post bounce phase of a core 
collapse supernova with $\numu/\nutau(\anumu/\anutau)$ having average energies 
more than that of $\nue(\anue)$. 
They can be detected by the new earth bound detector like 
SNO and Super-Kamiokande which are sensitive to neutrinos of all three flavours. 
In this letter we consider the effect of flavour oscillations on the neutrino 
flux and their expected number of events at the detector. We do a 
three-generation analysis and for the mass and mixing schemes we first 
consider the threefold maximal mixing model consistent with the solar 
and the atmospheric neutrino data and next a scenario with one 
$\Delta m^2 \sim 10^{-11} eV^2$ (solar range) and the other 
$\Delta m^2 \sim 10^{-18} eV^2$, for which the oscillation length is of 
the order of the supernova distance. 
In both these scenarios there are no matter effects in the 
resultant neutrino spectrum and one is concerned with vacuum oscillations. 
We find that though neutrino oscillations result in a depletion in the 
number of $\nue$ and $\anue$ coming from the supernova, the actual signals 
at the detectors are appreciable enhanced.}
\newpage 

The question whether neutrinos are massive or not has been answered. 
After 537 days of data on atmospheric neutrinos the Super-Kamiokande has 
finally confirmed the existence of non-zero oscillations and hence mass for 
the muon neutrinos \cite{skatm}. The Super-Kamiokande (SK) data confirmed the 
depletion in the atmospheric muon neutrino flux which the Kamiokande,  
IMB and the Soudan experiments had observed before. 
At 90\% C.L. the mixing parameters allowed by all the atmospheric 
neutrino experiments combined are \cite{garcia} 
$5\times 10^{-4} eV^2\leq \Delta m^2 \leq 6\times 10^{-3} eV^2$, 
$sin^2 2\theta \geq 0.8$ for the $\numu-\nutau$ channel and 
$10^{-3} \leq \Delta m^2 \leq 7\times 10^{-3} eV^2$, $sin^2 2\theta \geq 0.8$ 
for the $\numu-\nu_s$ ($\Delta m^2 > 0$) oscillation mode, 
while the oscillation parameter 
region for the $\numu-\nue$ mode is completely 
ruled out by the data from the CHOOZ experiment \cite{chooz}. 
The other puzzle 
which has warranted neutrino oscillations as a possible solution is the 
solar neutrino deficit problem. The three solar neutrino detectors, 
the Homestake, Kamiokande and Gallex (also Sage) 
have been observing 
neutrino flux far less than that predicted by the standard solar model 
\cite{bp}. This deficit can be explained by neutrino oscillations in vacuum 
for $\delta m^2 \sim 0.615\times 10^{-10} eV^2$ and $sin^2 2\theta \sim$ 
0.864 \cite{vac} or by MSW resonant flavour conversions \cite{msw} for 
$\Delta m^2 \sim 5.4\times 10^{-6} eV^2$ and $sin^2 2\theta \sim 7.9\times 
10^{-3}$(non-adiabatic solution) and $\Delta m^2 \sim 1.7\times 10^{-5} eV^2$ 
and $sin^2 2\theta \sim 0.69$(large angle solution) \cite{msw2}. The first 
results from the SK solar neutrino flux measurements favor the long 
wavelength vacuum oscillation solution with large angle mixing \cite{sksolar}. 
When the SK solar $\nu$ data is combined with the earlier data then 
the corresponding values are $\Delta m^2 \sim 6.5\times 10^{-11} eV^2, 
sin^2 2\theta \sim 0.75$ for vacuum oscillations and $\Delta m^2 
\sim 5\times 10^{-6} eV^2, sin^2 2\theta \sim 5.5\times 10^{-3}$ for 
non-adiabatic MSW resonant flavour conversion \cite{bahcall}.  
The threefold maximal mixing model \cite{smirnov,kimkim} 
can explain both the solar and the atmospheric neutrino data 
simultaneously \cite{har}. 
In a recent paper \cite{foot} it has been shown that the maximal 
mixing model can account for both the SK atmospheric data 
and the CHOOZ data provided the relevant $\Delta m^2$ is in the 
range $4\times 10^{-4}eV^2 \leq \Delta m^2 \leq 1.5\times 10^{-3}eV^2$, 
along with the solar $\nu$ data.  
One hopes to find a definite solution to the solar neutrino problem 
once the Sudbury Neutrino observatory (SNO) which is the first heavy water 
detector becomes operational \cite{sno}.
Though one of the principal motivation for these two detectors was to throw 
light on the solar neutrino problem but they are equally useful  
for detecting the neutrinos from a nearby supernova event.

The core of a massive star $(M\ge 8M_\odot)$ starts collapsing 
once it runs out of nuclear fuel.
The collapse continues to densities beyond the nuclear
matter density after which a bouncing of the in falling matter takes place
leading to supernova explosion and the formation of a protoneutron star.
Only a small fraction of the huge gravitational energy released in the process
goes into the explosion and all the rest of the energy is carried away by
neutrinos and antineutrinos of all three flavours. These neutrinos for
galactic supernova events can be detected by detectors like the SNO
and SK. In contrast to the solar, the atmospheric  
as well as the accelerator/
reactor neutrinos where one has neutrino flux of a single flavour at the
source, postbounce supernova neutrinos (antineutrinos) 
start from the source in
all three flavours but with $\nu_{\mu}/\nu_{\tau}$ 
($\anumu/\anutau$) having average energies
more than that of $\nu_e (\anue)$ and it is an interesting 
problem to study whether
their flux and their signal at the terrestrial $\nu$ detectors 
get appreciably altered in reaching the earth if 
neutrinos do oscillate.
In this work we give quantitative predictions for 
the number of neutrino events coming from a typical 
type II supernova at a distance of 10kpc in both SNO and SK and 
show how the number of events for each detection process would change 
in case oscillations do take place. 

There have been various attempts before to estimate the effect of 
non-zero neutrino mass and mixing on the expected neutrino signal 
from a galactic supernova. 
Matter enhanced resonant flavour conversion has been observed to have 
a large effect on the $\nue$ signal \cite{akh,bkg,qf}. The $\anue$ 
events of course remain unchanged in this case. With vacuum 
oscillations we can expect an increase in both the $\nue$ and $\anue$ 
signal.  
Burrows {\it et al.} \cite{bkg} have considered for SNO, the effect of 
vacuum oscillations as well and have found that 
with two-flavours the effect of vacuum oscillations on the signal is 
small, using their model predictions for the 
different $\nu$ luminosities. 

We have considered a three-generation mixing scheme and have 
calculated the effect of neutrino oscillations on the signal 
from a 20 $M_\odot$ supernova model developed recently \cite{totani}. 
First we do our calculations for the threefold maximal mixing model 
consistent with the solar ($\Delta m^2 \sim 10^{-11} eV^2$) and the 
atmospheric neutrino data ($\Delta m^2 \sim 10^{-3} eV^2$). 
For $\Delta m^2 \sim 10^{-3} eV^2$ 
normally we expect matter enhanced resonance in the supernova. But 
for the particular case of maximal mixing it has been shown before, 
both numerically \cite{harrison} and analytically \cite{bilenky}, that 
there are absolutely no matter effects in the resultant neutrino spectrum 
on earth. Though the arguments in both these previous papers are for 
solar neutrinos, extension to the case of supernova neutrinos is 
straightforward. Hence in this scheme we are concerned with vacuum 
oscillations. We also consider a second 
scenario where we take one of the mass square differences 
in the solar vacuum oscillation solution range 
while the other is 
$\sim 10^{-18} eV^2$ where the oscillations wavelength 
$\lambda \sim L$, the distance of the supernova from the earth and hence 
oscillations are observable in the neutrino spectrum. In this scenario 
of course there is no chance of a MSW resonance and we have vacuum 
oscillations. We find 
appreciable enhancement in the expected $\nue$ and $\anue$ charge 
current events for both SNO and 
SK in both scenarios even with vacuum oscillations. 
 
The differential number of neutrino events at the detector for a 
given reaction process is 
\begin{equation}
\frac{d^2 S_\nu}{dEdt}=\frac{n}{4\pi L^2} N_\nu(t) \sigma (E) f_\nu (E) 
\label{sig}
\end{equation}
One uses for the number of neutrinos produced at the source 
$N_\nu(t) = L_\nu(t)/\langle E_\nu(t)\rangle$ 
where $L_\nu(t)$ is the neutrino luminosity and 
$\langle E_\nu(t) \rangle$ is the average energy. In (\ref{sig}) 
$\sigma (E)$ is the reaction cross-section for the neutrino with 
the target particle, $L$ is the distance of the neutrino source 
from the detector(10kpc), 
$n$ is the number of detector particles for the reaction considered  
and $f_\nu (E)$ is the energy spectrum for the neutrino species involved. 
For the neutrino luminosity and average energy we use the values of 
Totani {\it et al.} \cite{totani} for a 20 $M_\odot$ type II supernova 
model based on the hydrodynamic code developed by Wilson and Mayle.  
Though in their paper Totani {\it et al.} 
observe that the neutrino spectrum is not a pure black body, but we 
as a first approximation use a Fermi-Dirac spectrum for the neutrinos, 
charaterised by the $\nu$ temperature alone 
for simplicity. The effect of a 
chemical potential is to cut the high energy tail of the neutrino spectrum 
and we also study it's effect on the the $\nu$ signal and on the  
enhancement of the signal when oscillations are introduced. We find the 
$\nu$ signal for the various detection processes as a function of 
energy by integrating out time from (\ref{sig}). By integrating (\ref{sig}) 
over energy as well we get the total number of events for the reaction 
concerned. These are the expected number of events.

In the presence of oscillations of massive neutrinos more 
energetic $\numu(\anumu)$ and $\nutau(\anutau)$ get transformed into $\nue
(\anue)$ which modifies the numbers that we obtain using (\ref{sig}). 
The general expression for the probability that an initial $\nu_\alpha$ gets 
converted to a $\nu_\beta$ after traveling a distance $L$ in vacuum is
\begin{equation}
P_{\nu_\alpha\nu_\beta}=\delta_\alpha \beta -4\sum_{j>i} 
U_{\alpha i}U_{\beta i}U_{\alpha j}U_{\beta j}sin^2 \frac{\pi L}
{\lambda_{ij}}
\label{prb}
\end{equation}
where $\alpha = e, \mu, \tau,..$ and $i,j = 1, 2, 3,..$
\begin{itemize}
\item $\lambda_{ij}=2.5\times 10^{-3}km \frac{E}{MeV}
\frac{eV^2}{\Delta m_{ij}^2}$\\
\item $\Delta m_{ij}^2=m_j^2 - m_i^2$\\
\end{itemize}
$U_{\alpha i}$ are the components of the mixing matrix. 
For the mass and mixing parameters we consider two scenarios. 
\begin{itemize}
\item {\bf scenario 1:} Here we consider threefold maximally mixed 
neutrinos with the mass spectrum $\Delta m_{13}^2 \approx 
\Delta m_{23}^2 \sim 10^{-3} eV^2$ corresponding to the atmospheric 
range while $\Delta m_{12}^2 \sim 10^{-11} eV^2$ in accordance with 
the solar neutrino problem.  
The oscillations due to all the mass differences are averaged out to 
1/2 as $\lambda << L$, and hence  
the expression for the various probabilities in this case relevant for 
us are \cite{kimkim,har}
\begin{equation}
P_{\nue \nue}=\frac{1}{3}
\label{prb1}
\end{equation}
\begin{equation}
P_{\numu \nue}+P_{\nutau \nue}=1-P_{\nue \nue}
\end{equation}
We call this Case 1. 

\item {\bf scenario 2:} Here we set $\Delta m_{12}^2 \sim 10^{-18} eV^2$ 
for which $\lambda \sim L$ and the oscillation effects are observable 
while $\Delta m_{13}^2 \approx \Delta m_{23}^2  \sim 10^{-11} eV^2$ 
(solar range). If we consider the Maiani parametrisation of the mixing 
matrix U \cite{giunti} then the expression for the probabilities are
\begin{equation}
P_{\nue \nue}=1-sin^2 2\theta_{12}cos^4 \theta_{13}sin^2 
\frac{\pi L}{\lambda_{12}} - \frac{1}{2} sin^2 2\theta_{13}
\label{prb2}
\end{equation}
\begin{equation}
P_{\numu \nue}+P_{\nutau \nue}=1-P_{\nue \nue}
\end{equation}
For this case the oscillations due to $\Delta m_{13}^2$ 
and $\Delta m_{23}^2$ are averaged out as the neutrinos 
travel to earth but those due to $\Delta m_{12}^2$ survive. 
For $\theta_{13}$ we consider two sets of 
values allowed by the solar $\nu$ data. We have done our calculations 
for $\sin^2 2\theta_{13}$ = 1.0 (the maximum allowed value) and with 
$\sin^2 2\theta_{13}$ = 0.75 (the best fit value) \cite{bahcall}. The 
first set is called Case 2a while the second is called Case 2b. 
Since nothing constrains $\Delta m_{12}^2$ in this scenario 
we can vary $\theta_{12}$ and study it's effect on the $\nu$ signal. 
We have tabulated our results for $\sin^2 2\theta_{12}$ = 1.0 since 
it gives the maximum increase in the signal from the no oscillation 
value. 
\end{itemize}
The corresponding expressions for the antineutrinos will be identical. 
We note that because the energy spectra of the $\numu$ and $\nutau$ 
are identical, we do not need to distinguish them and keep the 
combination 
$P_{\numu \nue}+P_{\nutau \nue}$. We have made here a three-generation 
analysis where all the three neutrino flavours are active. Hence if both 
the solar $\nu$ problem and the atmospheric $\nu$ anomaly require $\nu$ 
oscillation solutions, then in the {\bf scenario 2}, 
the atmospheric data has to be reproduced by $\numu-\nu_s$ 
oscillations. We are interested in this scenario as only 
with neutrinos from a supernova can one probe very small 
mass square differences $\sim 10^{-18} eV^2$. 
To find the number of events with oscillations we will have to fold the 
expression (\ref{sig}) with the expressions for survival and 
transition probabilities for the neutrinos 
for all the cases considered.

In Table 1 we report the calculated number of expected events for the 
main reactions in $\rm H_2O$ and $\rm D_2O$. 
Column 2 of Table 1 
gives the expected numbers for the model under consideration when 
the neutrino masses are assumed to be zero. Column 3,4,5 give the 
corresponding numbers for the two neutrino mixing scenarios 
that we have considered (see Table 1 for details). All the numbers 
tabulated have been calculated for 1 kton of detector mass. To get 
the actual numbers we have to multiply these numbers with the 
relevant fiducial mass of the detector. The efficiency of both the 
detectors (SNO and SK) is taken to be 1 \cite{totani,bevog,bevog2}. 
The energy threshold is taken to be 5 MeV for both SK \cite{bevog} and 
SNO \cite{bevog2}. For the cross-section of the $(\nue-d), (\anue-d), 
(\nu_x-d)$ 
and $(\anue-p)$ reactions we refer to \cite{burrows}. 
The cross-section of the $(\nue(\anue)-e^-)$ and $(\nu_x-e^-)$  
scattering has been taken from \cite{kolb} while the neutral current 
$(\nu_x-^{16}O)$ scattering cross-section is taken from \cite{bevog}. 
For the 
$^{16}O(\nue-e^-)^{16}F$ and $^{16}(\anue,e^+)^{16}N$ reactions we 
refer to \cite{haxton} where we have used the cross-sections 
for the detector with perfect efficiency.  
From a comparison of the 
predicted numbers in Table 1, it is evident that neutrino oscillations 
play a significant role in supernova neutrino detection. 
For the neutral current sector the number of 
events remain unchanged as the interaction is flavour blind.

The 32 kton of pure water in SK detects neutrinos 
primarily through the capture of $\anue$ on 
protons ($\anue p \rightarrow n e^+$) and $(\nue(\anue)-e^-)$ scattering. 
The energy threshold for $^{16}O(\nue,e^-)^{16}F$ is 15.4 MeV and 
that for $^{16}O(\anue,e^+)^{16}N$ is 11.4 MeV, hence these reactions   
are important only for very high energy neutrinos. 
The typical average energies of $\nue$ and $\anue$ from a type II 
supernova is 
about 11 MeV and 16 MeV respectively, so we do not expect significant 
contribution from these two reactions. This is evident from Table 1 
where the $^{16}O$ events are only 2.1\% of the total charge 
current signal at SK. 
As a result of mixing the mu and tau neutrinos 
and antineutrinos oscillate (with average energy $\sim$ 25 MeV) into 
$\nue$ and $\anue$ during their flight from the galactic 
supernova to the 
detector resulting in higher energy $\nue$ and $\anue$ 
and the number of $^{16}O$ 
events are increased appreciably (for Case 1 $(\nue-^{16}O)$ events go 
up by 13 times) so that after oscillations 
they are 7\% (Case 1) of the 
total charge current events at SK. 
The effect of oscillations on the ($\anue$-p) capture is to enhance 
the expected signal by about 25\% (Case 1).  
In all previous studies where the effect 
of MSW transition on the neutrino signal has been studied \cite{akh,qf}, 
there is no enhancement in the number of expected events for the ($\anue$-p) 
sector while we do get a significant change in the expected 
signal with vacuum oscillations. 
For the ($\nue(\anue) - e^-$) scattering the effect of 
oscillation is very small. 

The SNO is the world's first heavy water detector made of 1 kton of 
pure $\rm D_2O$ surrounded by ultra pure $\rm H_2O$. 
There are $10^4$ phototubes around this entire volume which can 
view only the inner 1.4 kton of water efficiently \cite{bevog2}. 
We find about 99\% increase in $(\nue-d)$ events and about 46\% increase 
in $(\anue-d)$ events for the Case 1. 
From the column 2 of Table 1 
we can see that there are more $(\anue-d)$ than $(\nue-d)$ 
events even though there are more $\nue$ than $\anue$ coming 
from the supernova. This is  because the reaction cross-section $\sigma \sim 
E^{2.3}$ and the $\anue$ spectrum is harder than the $\nue$ 
spectrum. This also results in a greater enhancement due to oscillations 
for the $(\nue-d)$ events, as the difference between the energies of the 
$\nue$ and $\numu(\nutau)$ is greater than those between $\anue$ and 
$\anumu(\anutau)$ and hence the effect on the $\nue$ events is more. 
As a result after oscillations are switched on the number of $(\nue-d)$ 
events supersede the $(\anue-d)$ events. 
We observe a similar effect for the $^{16}O$ events, where the $\anue$ 
signal without oscillations is more than the $\nue$ signal, 
while the effect of oscillations 
is more for the latter. The effect is more magnified in this case due to 
the very strong energy dependence of the reaction cross-section and 
also due to the fact that the energy threshold for $(\anue-^{16}O)$ 
event is lower than for the $(\nue-^{16}O)$ event. 
In Fig. 1 we plot the signal due to the ($\nue-d$) events as a 
function of energy , 
without oscillations and with oscillations for the Case 1 and 
Case 2b. All the features mentioned are clearly seen. The plot 
for the Case 2b clearly shows oscillations. 

In Fig. 2 we plot the cumulative fluence of the $\nue$ 
coming from the supernova at 10 kpc without oscillations and with 
oscillations for Case 1 and Case 2b. It is seen that the result of 
oscillation in fact is to reduce the total number of $\nue$. Yet 
as seen from Table 1, we have 
obtained significant increase in the $(\nue-d)$ events and 
the $(\nue -^{16}O)$ events. The solution to 
this apparent anomaly lies in the fact that the 
cross-section of these reactions 
are strongly energy dependent. As a result of oscillations the $\nue$ flux 
though depleted in number, gets enriched in high energy neutrinos. 
It is these higher energy neutrinos which enhance the $\nu$ signal at the 
detector. This also explains the difference in the degree of enhancement 
for the different processes. For the $(\nue-d)$ and $(\nue - ^{16}O)$ 
events, especially for the latter, the effect is huge while for the 
$(\nue-e^-)$ scattering it is negligible as it's reaction cross-section 
is only linearly proportional to E.  
Due to their high energy dependent $\sigma$ 
the $^{16}O(\nue,e^-)^{16}F$ events turn out to be 
extremely sensitive to oscillations. 
A similar argument holds true for the case of the antineutrinos, only 
here the effect of oscillations is less than in the case for the 
neutrinos as the difference between the energies of the $\anue$ and 
$\anumu/\anutau$ is comparatively less as discussed earlier. 

For the {\bf scenario 2} we have studied the effect of the mixing 
angles on the signal. For a fixed $\theta_{13}$ the effect of 
oscillations is enhanced if we raise $\theta_{12}$. The effect of 
$\theta_{13}$ is more subtle. The effect of oscillations increase 
with $\theta_{13}$ initially and then 
decrease. 
We have also checked the effect of a chemical potential 
$\mu$ on the neutrino signal. A non-zero $\mu$ cuts the high energy 
tail of the neutrino signal as a result of which the total signal goes 
down for both with and without oscillations, the effect being greater 
for the more energy sensitive reactions. 

With the supernova model of Totani {\it et al.} \cite{totani}, we 
have obtained oscillation effects in the expected $\nu$ signal which 
are significantly larger than those obtained by Burrows {\it et al.} 
\cite{bkg,bkg1}. In the model that Burrows {\it et al.} use in their 
study, the $\nu$ luminosities $L_\nu$ are more than those for 
Totani {\it et al.} model, but the average energy is much smaller, 
particularly for the $\anue$ and $\nu_{\mu,\tau}(\bar\nu_{\mu,\tau})$. 
Hence their $\numu$ spectra lacks in high energy neutrinos which 
results in almost negligible effect of oscillations in their case. 
Again in the model of Burrows {\it et al.} 
the average energies decrease with time while in the model of Totani 
{\it et al.} not only the average energies but also the difference between 
the average energies of $\nue(\anue)$ and $\nu_{\mu,\tau}(\bar\nu_{\mu,\tau})$ 
increases with time. The effect of all these is to magnify the effect of 
oscillations in our case. 

In conclusion, we have shown that with the model of Totani {\it et al.} 
even with vacuum oscillations we obtain appreciable enhancement in 
the expected $\nu$ signal in SNO and SK even though the number of 
neutrinos arriving at the detector from the supernova goes down. In 
contrast to the case where we have MSW resonance in the supernova, with 
vacuum oscillations we get enhancement for both $\nue$ as well as $\anue$ 
events. If we have a galactic supernova event in the near future and if we 
get a distortion in the neutrino spectrum and an enhancement in the signal, 
for both $\nue$ as well as $\anue$ then that would indicate vacuum 
neutrino oscillations.

{\small The authors wish to thank S.Goswami, A.Raychaudhuri and A.Ray 
for useful discussions and J.Beacom for valuable suggestions. The work 
of D.M. and K.K. is partially supported by the Eastern Centre for Reasearch 
in Astrophysics, India}

\newpage

\begin{description}
\item{{\bf Table 1}} The expected number of neutrino events for a 1 kton  
water cerenkov detector 
\end{description}
\[
\begin{tabular}{|c|c|c|c|c|} \hline 
{} & {\rm signal}
& \multicolumn{3}{c|}{\rm signal\hspace{2mm} with \hspace{2mm}oscillation}\\ 
\cline{3-5} 
{\rm reaction} & {\rm without} & 
{\rm {\bf scenario\hspace{1mm}1}} & 
\multicolumn{2}{c|}{\rm {\bf scenario\hspace{1mm}2}}\\ \cline{3-5} 
{} & {\rm oscillation} & {Case 1}  
& {\rm Case2a} & {\rm Case2b} 
\\ \hline
{$\nu_e+d\rightarrow p+p+e^-$} & {78} & {155} 
& {150} & {153}\\ \hline
{$\bar\nu_e +d\rightarrow n+n+e^+$} & {93} & {136} 
& {133} & {135}\\ \hline
{$\nu_x+d\rightarrow n+p+\nu_x$} & {455} & {455} & {455} & 
{455} \\ \hline
{$\bar\nu_e +p\rightarrow n+e^+$} & {263} & {330} 
& {326} & {329}\\ \hline
{$\nu_e +e^- \rightarrow \nu_e +e^-$} 
& {4.68} & {5.68} & {5.61} & {5.66} \\ \hline
{$\bar\nu_e+e^- \rightarrow \bar\nu_e+e^-$} 
& {1.54} & {1.77} & {1.76} & {1.77} \\ \hline
{$\nu_{\mu,\tau}(\bar\nu_{\mu,\tau}) +e^- \rightarrow 
\nu_{\mu,\tau}(\bar\nu_{\mu,\tau}) +e^-$} 
& {3.87} & {3.55} & {3.50} & {3.53} \\ \hline
{$\nu_e +^{16}O \rightarrow e^- +^{16}F$} & {1.13} & {14.58} 
& {13.78} & {14.45} \\ \hline
{$\bar\nu_e + ^{16}O\rightarrow e^+ + ^{16}N$} & {4.57} & {10.62} 
& {10.23} & {10.53} \\ \hline
{$\nu_x +^{16}O \rightarrow \nu_x +\gamma +X$} & {13.6} & {13.6} & 
{13.6} & {13.6} \\ \hline 
\end{tabular}
\]

\newpage
\begin{center}
{\bf Figure Captions}
\end{center}

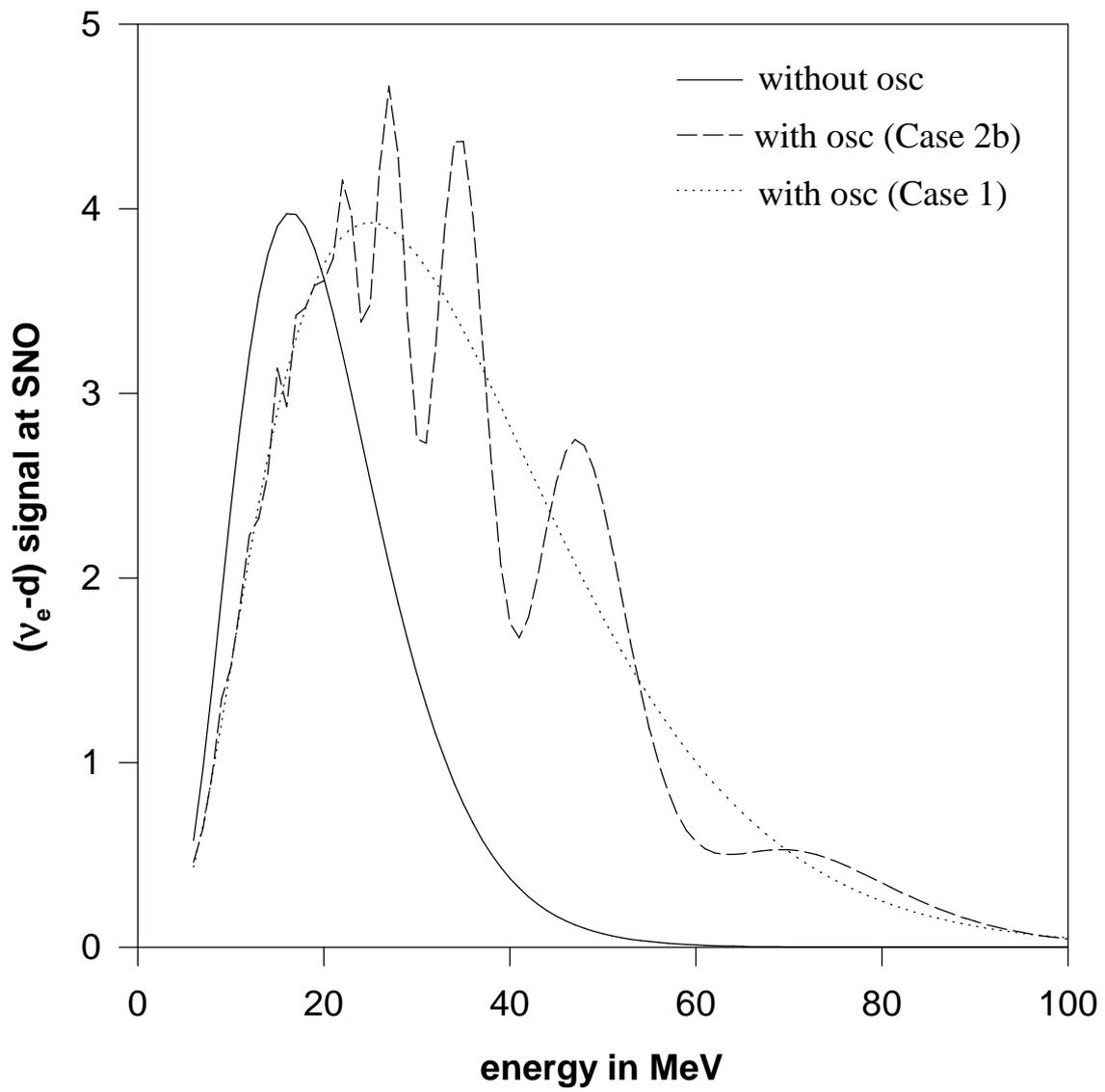
{\bf Fig. 1} The ($\nue-d$) signal at SNO vs neutrino energy 
without and with oscillations for 
the Case 1 and Case 2b

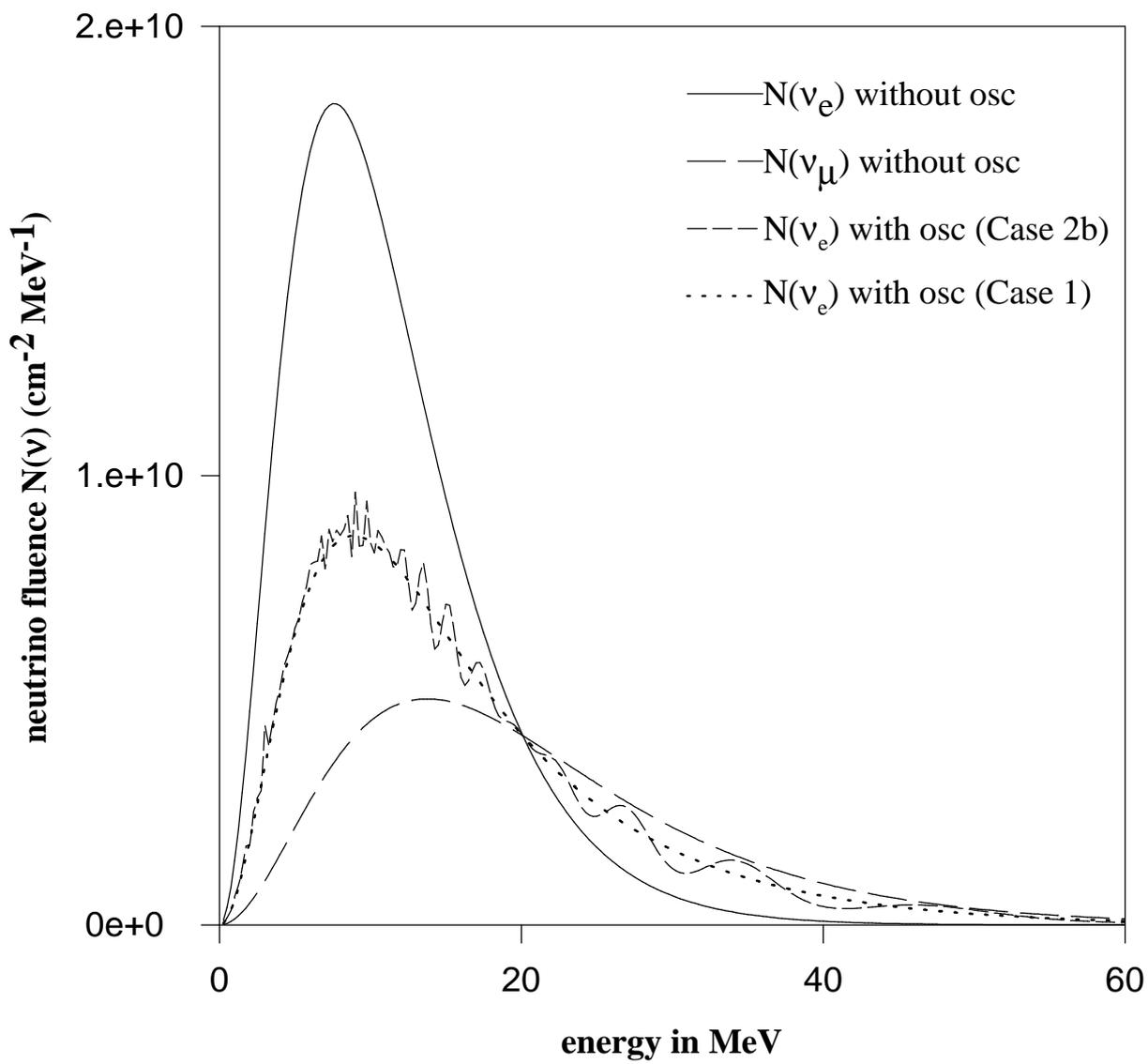
{\bf Fig. 2} The cumulative $\nue$ fluence as a function of the neutrino 
energy without and with oscillations for the Case 1 and Case 2b. Also 
shown is the $\numu$ fluence for comparison.

\end{document}